\documentclass[twocolumn,showpacs,preprintnumbers,amsmath,amssymb]{revtex4}

\usepackage{bm}
\usepackage[colorlinks=true,linkcolor=blue,citecolor=blue,urlcolor=blue]{hyperref}
\usepackage{color}

\usepackage{graphicx}
\usepackage{epstopdf}

\begin{document}


\title{Simple top-down preparation of magnetic Bi$_{0.9}$Gd$_{0.1}$Fe$_{1-x}$Ti$_x$O$_3$ nanoparticles by ultrasonication of multiferroic bulk material}

\author{M. A. Basith}
\email[Author to whom correspondence should be addressed (e-mail): ]{mabasith@phy.buet.ac.bd}
\author{A. Quader, M. A. Rahman}
\affiliation{Department of Physics, Bangladesh University of Engineering and Technology, Dhaka-1000, Bangladesh.
}
\author{D.-T. Ngo}
\email[Author to whom correspondence should be addressed (e-mail): ]{dngo@nanotech.dtu.dk}
\author{K. M{\o}lhave}
\affiliation{Department of Micro-and Nanotechnology, Technical University of Denmark, Kgs. Lyngby 2800, Denmark.}

\author{B. L. Sinha}
\affiliation{Department of Science and Humanities, MIST, Dhaka, Bangladesh.
}
\author{Bashir Ahmmad}
\email[Author to whom correspondence should be addressed (e-mail): ]{arima@yz.yamagata-u.ac.jp}
\author{Fumihiko Hirose}
\affiliation{Graduate School of Science and Engineering, Yamagata University, 4-3-16 Jonan, Yonezawa 992-8510, Japan.}





\begin{abstract}
We present a simple technique to synthesize ultrafine nanoparticles directly from bulk multiferroic perovskite powder. The starting materials, which were ceramic pellets of the nominal compositions of Bi$_{0.9}$Gd$_{0.1}$Fe$_{1-x}$Ti$_x$O$_3$ (x = 0.00-0.20), were prepared initially by a solid state reaction technique, then ground into micrometer-sized powders and mixed with isopropanol or water in an ultrasonic bath. The particle size was studied as a function of sonication time with transmission electron microscopic imaging and electron diffraction that confirmed the formation of a large fraction of single-crystalline nanoparticles with a mean size of 11-13 nm. A significant improvement in the magnetic behavior of  Bi$_{0.9}$Gd$_{0.1}$Fe$_{1-x}$Ti$_x$O$_3$ nanoparticles compared to their bulk counterparts was observed at room temperature. This sonication technique may be considered as a simple and promising route to prepare ultrafine nanoparticles for functional applications.  
   
\end{abstract}

\maketitle
\section{Introduction} \label{I}
Multiferroic materials exhibit simultaneous presence of (anti)ferroelectricity, (anti)ferromagnetism, and/or ferroelasticity in the same phase \cite{ref1,ref2,ref3,ref53, ref77}. The combination of 'ferro'-orders in multiferroics often means that the different properties interact with each other. This then allows the possibility that one can switch magnetically ordered states using electric fields or electrically ordered states using magnetic fields. These materials have attracted considerable research interest due to their potential applications in data storage, emerging field of spintronics, switchable spin valves, high frequency filters and sensors \cite{ref6,ref7}. Among the limited choices offered by the multiferroic materials, BiFeO$_3$ with rhombohedrally distorted perovskite structure is the most promising since it can exhibit multifunctional activities at room temperature (RT) \cite{ref6,ref7}. 
However, pure phase of BiFeO$_3$ is difficult to obtain \cite{ref8,ref9} and various impurity phases of bulk BiFeO$_3$ have been reported, mainly comprising of Bi$_2$Fe$_4$O$_9$, Bi$_{36}$Fe$_{24}$O$_{57}$ and Bi$_{25}$FeO$_{40}$ \cite{ref10,ref11}. Moreover, the bulk BiFeO$_3$ is characterized by serious current leakage problems due to the existence of a large number of charge centres caused by oxygen ion vacancies \cite{ref12, ref59, ref52}. Besides, in BiFeO$_3$, magnetic ordering is of antiferromagnetic type, having a spiral modulated spin structure (SMSS) with an incommensurate long-wavelength period of 62 nm \cite{ref12, ref52}. This spiral spin structure cancels the macroscopic magnetization and prevents the observation of the linear magnetoelectric effect \cite{ref13,ref14, ref15}.  These problems ultimately limit the use of bulk BiFeO$_3$ in functional applications.  It is evident that nanoparticles of BiFeO$_3$, especially those with a particle size on the order of or smaller than the 62 nm SMSS exhibit improved ferroelectric and magnetic properties \cite{ref53,ref16,ref17,ref51}. 

Synthesis of BiFeO$_3$ multiferroic nanoparticles hence requires a particle size of the order or smaller than 62 nm \cite{ref75}. To date, most of the published results reported the multiferroic properties of ceramics \cite{ref8,ref9} and thin film \cite{ref91} systems of BiFeO$_3$ and it is still a challenge to synthesize BiFeO$_3$ nanostructures \cite{ref7,ref75,ref78}. So far various chemistry based routes like the sol-gel method \cite{ref16, ref79}, electrospray method \cite{ref61}, the combustion synthesis process \cite{ref81}, sonochemical synthesis process \cite{ref63,ref94,ref96}, were applied to synthesize multiferroic nanoparticles. Most of these chemical methods for the synthesis of multiferroic nanoparticles are either based on complex solution processes or involve toxic precursors \cite{ref7}.

This paper presents a simple route to prepare ultrafine nanoparticles from multiferroic Bi$_{0.9}$Gd$_{0.1}$Fe$_{1-x}$Ti$_x$O$_3$ (x = 0.00, 0.10 and 0.20) ceramics by the application of ultrasonic energy in a process called sonofragmentation. Sonofragmentation has been used to create nanoparticle dispersions from bulk powder producing nanoparticle fractions \cite{ref82,ref83,ref84} and even inherently strong materials like carbon nanotubes can be fragmented by sonication \cite{ref85}. The sonofragmentation technique is in fact a one-step synthesis technique to produce nanoparticles directly from bulk powders and might have the additional advantage that the chemistry of the particles likely will not be altered and hence bulk and nanoparticle materials will be more directly comparable in terms of multiferroic properties than materials produced by different synthesis methods. In our investigation, the magnetic properties of Bi$_{0.9}$Gd$_{0.1}$Fe$_{1-x}$Ti$_x$O$_3$ (x = 0.00, 0.10 and 0.20)  nanoparticles synthesized by ultrasonication were investigated at room temperature and compared with their bulk counterparts. To estimate the concentration of oxygen vacancies in bulk polycrystalline samples as well as nanoparticles we have carried out X-ray photoelectron spectroscopy (XPS) analysis.




\section{Experimental details} \label{II}
Ceramic pellets of mutiferroics with nominal compositions of Bi$_{0.9}$Gd$_{0.1}$Fe$_{1-x}$Ti$_x$O$_3$ (x = 0.0, 0.1 and 0.2) were prepared by conventional solid state reaction technique. Details of the preparation process were described in our previous work published elsewhere \cite{ref18}. Scanning Electron Microscopy (SEM) imaging confirmed surface morphology of the pellets with micron-sized grains \cite{ref18}. The pellets were then ground into powder by performing manual grinding for 15 minutes. The obtained powders were subsequently mixed with isopropanol with a ratio of 50 mg powder and 10 ml isopropanol with a mass percentage of $\sim$ 0.5 \%. Then, the mixtures of isopropanol and powder were put into an ultrasonic bath. The sonication time was varied from 15 to 60 minutes. After 6 hours, $\sim$ 78 \% of the mass had precipitated and the supernatant was used for transmission electron microscopy (TEM) investigation. Morphology and physical structure of the nanoparticles were studied using a FEI TECNAI TEM with a 200 kV accelerated voltage. For magnetic measurements, the particles in the solution were dried naturally to a condensed pellet. The magnetic properties of the nanoparticles were measured using an alternating gradient force magnetometer (AGFM) with a maximum applied field of 20 kOe. The magnetic moment was calibrated using a Ni standard specimen. To estimate the concentration of oxygen vacancies X-ray photoelectron spectroscopy (XPS, ULVAC-PHI Inc., Model 1600) analysis was carried out with a Mg-K{$\alpha$} radiation source.

\section{Results and discussions} \label{III}

\begin{figure}[hh]
\centering
\includegraphics[width=8cm]{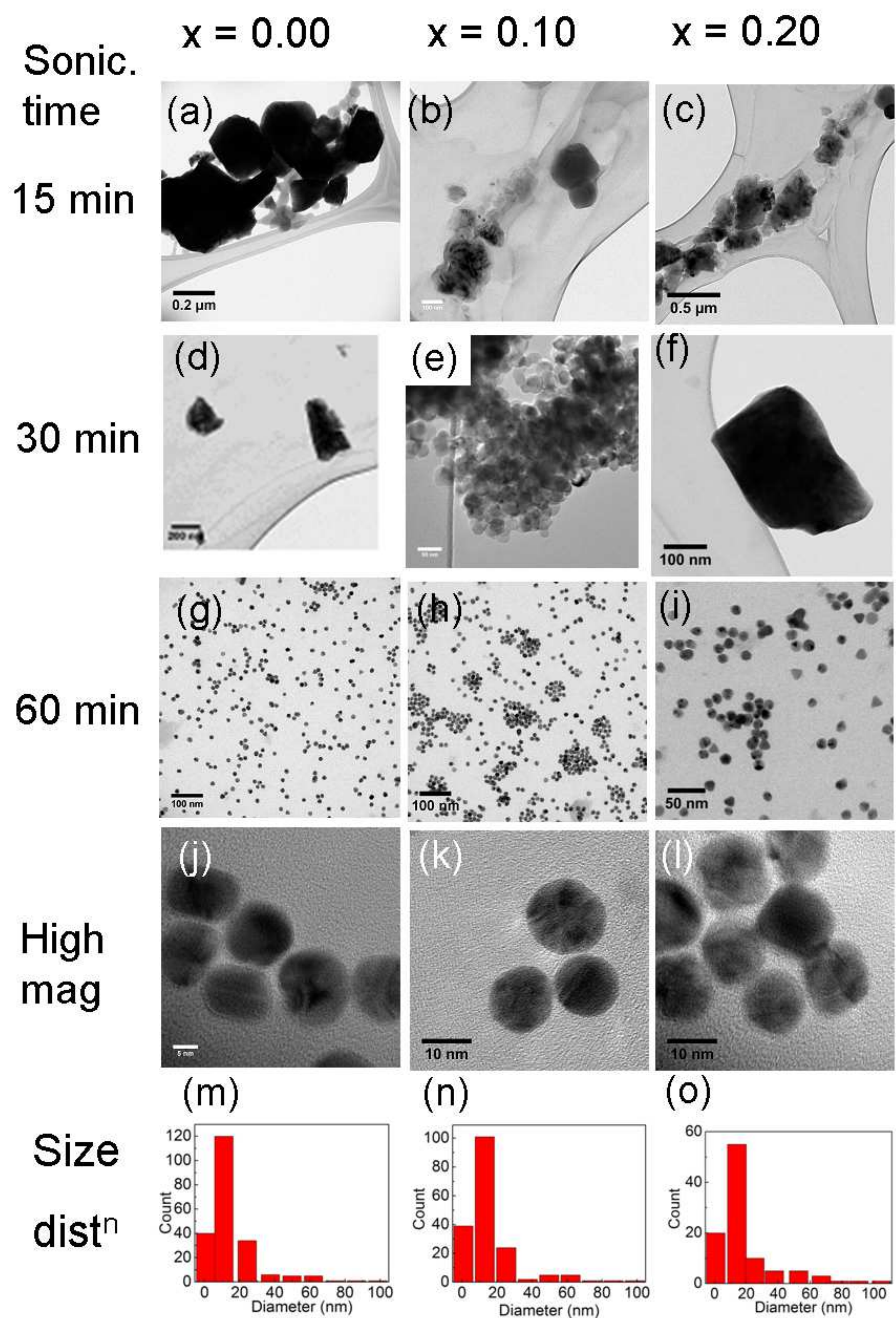}
\caption{BF TEM images of the Bi$_{0.9}$Gd$_{0.1}$Fe$_{1-x}$Ti$_x$O$_3$ (x = 0.00-0.20) particles obtained at various sonication time: (a-c) 15 minutes, (d-f)  30 minutes, (g-i) 60 minutes. (j, k and l) are the HRTEM images obtained at 60 mins of sonication. The plots (m, n and o) show distribution of nanoparticles deduced from frames (g, h and i), respectively.} \label{fig1}
\end{figure} 
Figure \ref{fig1} shows bright-field (BF) TEM images of Bi$_{0.9}$Gd$_{0.1}$Fe$_{1-x}$Ti$_x$O$_3$ (x = 0.00-0.20) particles obtained at various sonication time: (a-c) 15 minutes, (d-f)  30 minutes and (g-i) 60 minutes. The left, middle and right columns are for compositions x = 0.00, x= 0.10 and 0.20, respectively. Figures \ref{fig1} (a-c) present the state in which the manually-ground powder was mixed with isopropanol in ultrasound, and the large-grain size powders started to break into smaller grain with the aid of the mechanical energy from the ultrasonic wave. In a short time of sonication (15 minutes), the grains were just about to be broken into sub-micron particles, however, still condensed together like micrometer sized particles [Figs. \ref{fig1}(a-c)]. As soon as the time of sonication was increased to 30 minutes, the particles started to separate  from each other and thereby their sizes were decreased [Figs. \ref{fig1}(d-f)]. The sonication time was increased then to 60 minutes and this resulted in ultrafine and isolated nanoparticles with a very narrow distribution of average sizes around $11\pm 2$ nm for x = 0.00, $12\pm 2$ nm for x = 0.10, and $13\pm 2$ nm for x = 0.20 compositions as shown in TEM images Figs. \ref{fig1} (g-i), respectively. Figures \ref{fig1} (j, k and l) demonstrate high resolution (HR) TEM images of the Bi$_{0.9}$Gd$_{0.1}$Fe$_{1-x}$Ti$_x$O$_3$ (x = 0.00-0.20) nanoparticles obtained at 60 minutes of sonication. These HRTEM images [Figs. \ref{fig1} (j, k and l)] are a clear evidence of the ultrafine single crystal nanoparticles synthesized using a sonication time of 60 minutes. Figures \ref{fig1}(m, n and o) show the distribution of the synthesized nanoparticles deduced from images (g, h and i), respectively.

\begin{figure}[!hh]
\centering
\includegraphics[width=8cm]{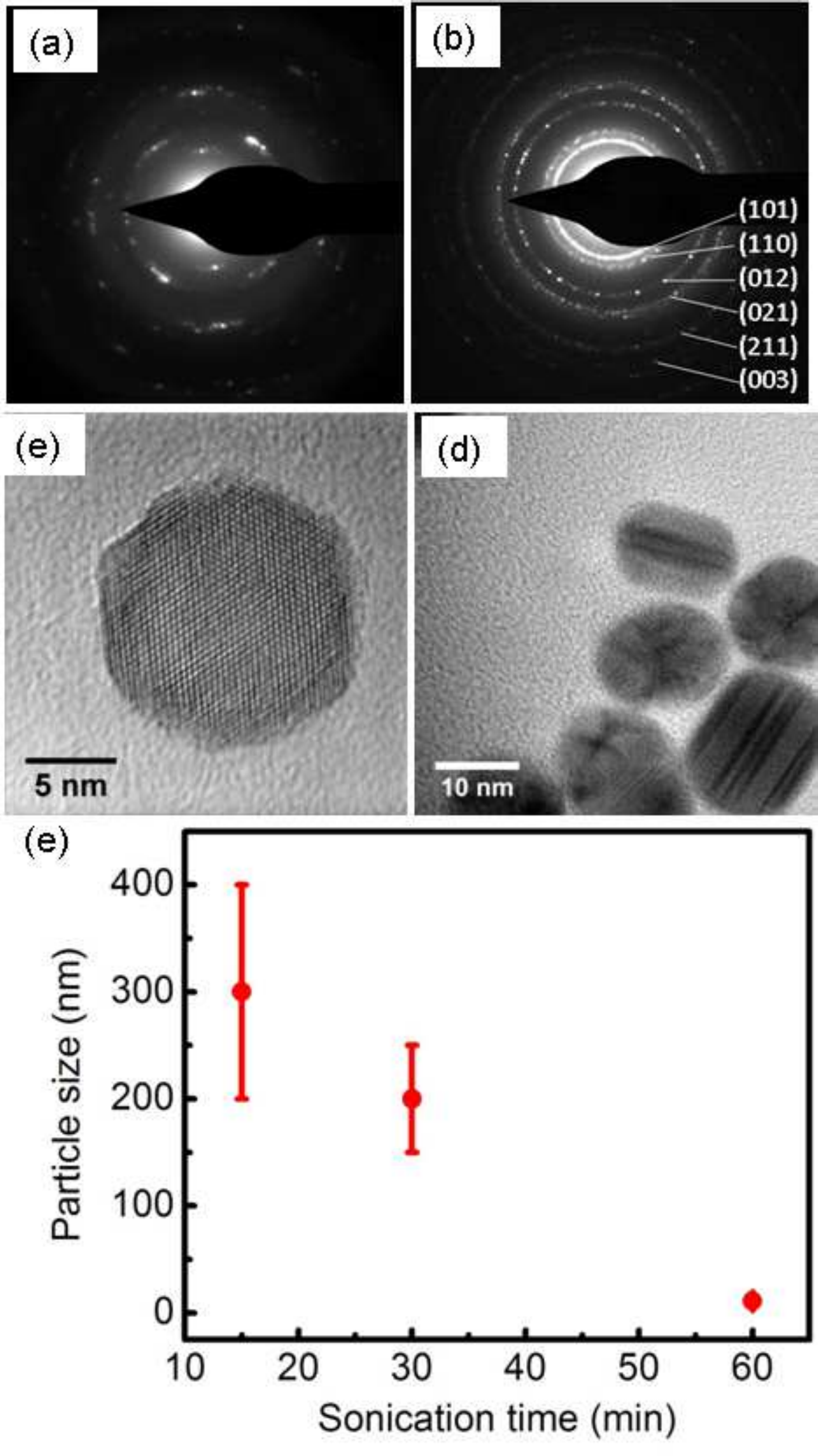}
\caption{ (a) The SAED patterns of the Bi$_{0.9}$Gd$_{0.1}$Fe$_{1-x}$Ti$_x$O$_3$ (x = 0.0) particles sonicated 30 minutes showing discontinuous rings. Detailed TEM investigation of the Bi$_{0.9}$Gd$_{0.1}$Fe$_{1-x}$Ti$_x$O$_3$ (x = 0.00) nanoparticle (first column in Fig. 1) sonicated at 60 minutes are given in images (b-d). (b) SAED of nanoparticles showing multiple peaks of  monocrystalline  patterns. (c) HRTEM showing monocrystalline particle. (d) HRTEM showing twin stacking in single particle. (e) The variation of particle size as a function of sonication time for a fixed sonication power.} \label{fig2}
\end{figure}

Figure \ref{fig2}(a) shows selected area electron diffraction (SAED) patterns of the Bi$_{0.9}$Gd$_{0.1}$Fe$_{1-x}$Ti$_x$O$_3$ with x = 0.00 particles (notably, x = 0.00 is the Ti undoped Bi$_{0.9}$Gd$_{0.1}$FeO$_3$ composition throughout the investigation) obtained after a sonication time of 30 minutes. This is the  state which confirms a rhombohedral \cite{ref51} structure of the isolated particles. The discontinuous diffraction rings also indicates single crystalline particles breaking up into smaller single crystalline nanoparticles. Figure \ref{fig2}(b) illustrates SAED of the Bi$_{0.9}$Gd$_{0.1}$Fe$_{1-x}$Ti$_x$O$_3$ (x = 0.00) nanoparticle (first column in Fig. 1) sonicated at 60 minutes and demonstrates clearly multiple peaks of  monocrystalline  patterns. The diffraction rings of SAED pattern in this state are well-defined and in good agreement with the rhombohedral structure of the nanoparticles with lattice constant of a = 0.56 nm, and $\alpha$ = 59.4 degree obtained by indexing the SAED pattern. These values are quite consistent with the recent results of the BiFeO$_3$ nanoparticles \cite{ref19, ref20} synthesized by wet chemical techniques. This indicates that the rhombohedral crystal structure of the bulk BiFeO$_3$ \cite{ref6,ref7} remained unaltered in these synthesized nanoparticles.  The HRTEM images demonstrate of a monocrystalline particle (Fig. \ref{fig2}(c)) and the existence of twin stacking faults (Fig. \ref{fig2}(d)) in the synthesized nanoparticles. These crystallographic stacking faults could also enhance the magnetic properties of the nanoparticles \cite{ref86,ref87, ref88}. The stacking faults were also reported to relax the strain in the nanoparticles \cite{ref86,ref87}, thus influence the magneoelastic energy. In some cases, this would help to enhance the anisotropy of the nanoparticles \cite{ref87,ref88}, especially in nanoparticles with ferromagnetic spin structure that we show below these particles have.

The effect of sonication time to breakdown micrometer sized particles was also demonstrated. Figure \ref{fig2}(e) shows the variation of the size of Bi$_{0.9}$Gd$_{0.1}$Fe$_{1-x}$Ti$_x$O$_3$ (x = 0.0) nanoparticle as a function of sonication time for a fixed power of the ultrasonic bath (50 W). We have observed clearly that a sonication time of 60 minutes was sufficient to produce monocrystalline nanoparticles. A further increment of the sonication time up to 90 minutes does not causes notable changes in the average size of the synthesized nanoparticles. The asymptotic behavior in Fig. \ref{fig2}(e)
indicates that within the studied time frame, sonication induced aggregation of the nanoparticles is not occurring \cite{ref97}.

The yield of nanoparticles was estimated by measuring the mass of precipiate and supernatant after leaving the sonicated dispersion to settle for 6 hours where 22 \% of the initial powder mass was converted into the supernatant fraction containing nanoparticles. Prior to these measurements the precipitate and supernatant were dried at 105$^o$C for 2.5 hours.  

The structural analysis of the Bi$_{0.9}$Gd$_{0.1}$Fe$_{1-x}$Ti$_x$O$_3$ (x = 0.10 and 0.20) nanoparticles sonicated at 60 minutes were also performed by capturing SAED patterns (images not shown here). Indexing the SAED patterns of the nanoparticles also ensured the rhombohedral structure of the Bi$_{0.9}$Gd$_{0.1}$Fe$_{0.9}$Ti$_{0.1}$O$_3$ and Bi$_{0.9}$Gd$_{0.1}$Fe$_{0.8}$Ti$_{0.2}$O$_3$ nanoparticles with the lattice constants of a = 0.55 nm, and 0.54 nm; and $\alpha$ = 59.3, and 59.3 $\deg$, respectively. The rhomboheral structure with reduced lattice constants due to the substitution of Ti for Fe is consistent with recent results \cite{ref18,ref19, ref20} on the doped BiFeO$_3$ multiferroics.

\begin{figure}[!hh]
\centering
\includegraphics[width=8 cm]{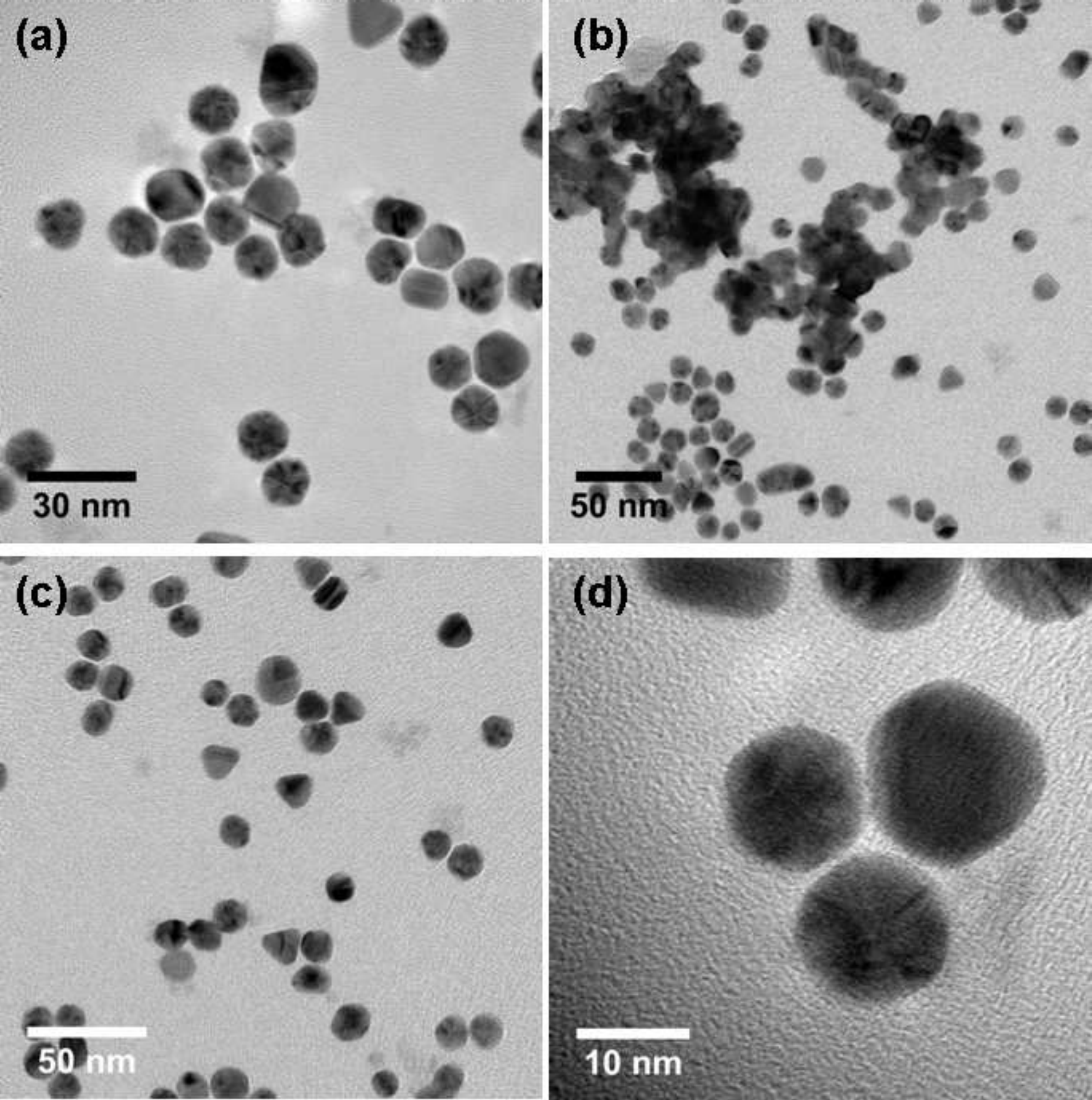}
\caption{TEM images of the Bi$_{0.9}$Gd$_{0.1}$Fe$_{1-x}$Ti$_x$O$_3$ nanoparticles obtained at 60 minutes of sonication in distilled water: (a) x = 0.00, (b) x = 0.10, (c) x = 0.20. The frame (d) is a HRTEM image of the nanoparticles in frame (c).} \label{fig3}
\end{figure}

It is interesting to notice that similar evolution also occurs when we use distilled water instead of isopropanol to mix the powder in ultrasonic cleaner. The TEM images of the Bi$_{0.9}$Gd$_{0.1}$Fe$_{1-x}$Ti$_x$O$_3$ (x = 0.00 - 0.20) nanoparticles sonicated for 60 minutes using distilled water  are shown in Figs. \ref{fig3} (a-c): (a) x = 0.00, (b) x = 0.10, (c) x = 0.20. The frame (d) is a HRTEM image of the nanoparticles in frame (c). The average size of the synthesized nanoparticles is $13\pm 2$ nm if we mix  the powders with distilled water instead of isopropanol. Notably, use of distilled water instead of isopropanol for the production of stable nanoparticles does not cause any difference in sonication time and the optimum sonication time is again 60 minutes.

\begin{figure}[hh]
\centering
\includegraphics[width=8cm]{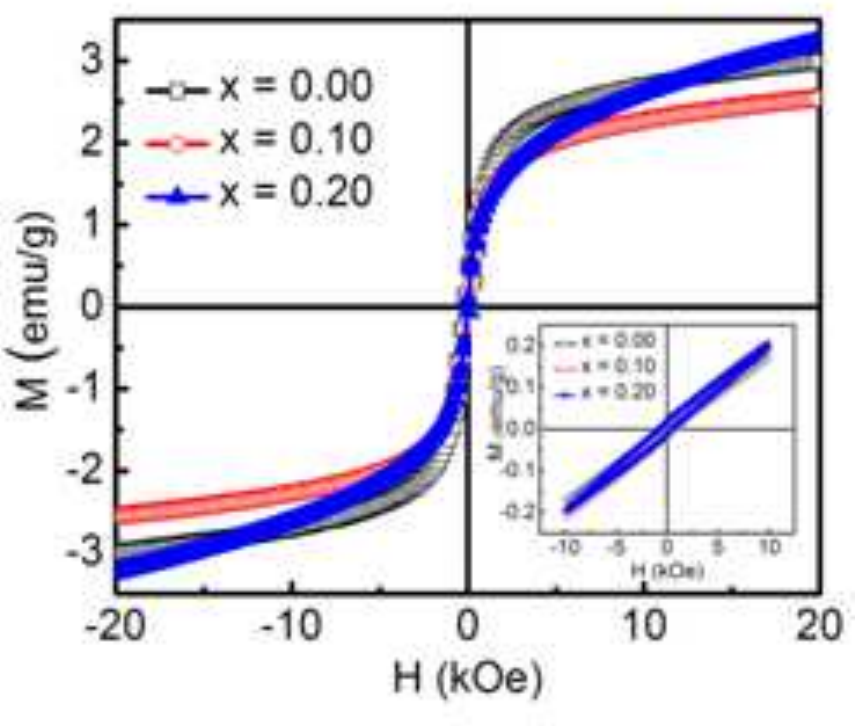}
\caption{Hysteresis loops of the Bi$_{0.9}$Gd$_{0.1}$Fe$_{1-x}$Ti$_x$O$_3$ (x = 0-0.20) nanoparticles obtained after a sonication time of 60 minutes. The inset shows hysteresis loops of the bulk samples.} \label{fig5}
\end{figure}

In the next stage of this investigation, magnetization versus magnetic field (M-H) hysteresis loops were carried out and results were displayed in Fig. \ref{fig5}. The inset of Fig. \ref{fig5} also demonstrates M-H hysteresis loops of the bulk polycrystalline ceramics Bi$_{0.9}$Gd$_{0.1}$Fe$_{1-x}$Ti$_x$O$_3$ (x = 0.00-0.20) \cite{ref18}. The obtained nanoparticles are expected to exhibit excellent properties, either ferromagnetics or ferroelectrics. Indeed, the nearly saturated M-H hysteresis loops of the synthesized nanoparticles clearly demonstrate improved magnetic properties compared to that of bulk ceramics [Fig. \ref{fig5}]. Notably, the bulk polycrystalline Bi$_{0.9}$Gd$_{0.1}$Fe$_{1-x}$Ti$_x$O$_3$ ceramics exhibit an almost linear relationship between the magnetic field and the magnetization \cite{ref18} as was shown in the inset of Fig. \ref{fig5}. This is the expected behaviour for an antiferromagnet below T$_{Neel}$. Compared to the bulk ceramic samples \cite{ref18}, these nanoparticles possess much higher saturation magnetization M$_s$. The M$_s$ value was determined from the intercept of two linear lines drawn through the low- and high-field regions of the M-H hysteresis loops \cite{ref55}. The calculated values of M$_s$ along with remanent magnetization (M$_r$) and coercivity (H$_c$) of Bi$_{0.9}$Gd$_{0.1}$Fe$_{1-x}$Ti$_x$O$_3$ (x = 0-0.20) nanoparticles are inserted in table 1. This table also displays the corresponding M$_s$, M$_r$ and H$_c$ values of the bulk Bi$_{0.9}$Gd$_{0.1}$Fe$_{1-x}$Ti$_x$O$_3$ ceramics observed in Ref. \cite{ref18}. Furthermore, the M-H hysteresis loops of the synthesized nanoparticles exhibit a single-phase-like magnetization behaviour compared to the linear curve of the bulk, and this ultimately indicate qualitatively that a large fraction of the ultrasonicated specimen was converted into nanoparticles from bulk powders.  

Our results can be compared to a recent study where single crystalline BiFeO$_3$ nanoparticles were synthesized using facile sol-gel methodology based on the glycol gel reaction \cite{ref16}. The highest M$_s$ for BiFeO$_3$ nanoparticles was 1.55 emu/g at 50 kOe with an average particle size of 14 nm \cite{ref16}. The M$_s$ of Gd doped multiferroic Bi$_{0.9}$Gd$_{0.1}$FeO$_3$ nanoparticles synthesized by polyol-mediated method \cite{ref51} with an average particle size of 34 nm  was 0.75 emu/g at 30 kOe \cite{ref51}. Notably, in this present investigation, the M$_s$ for Bi$_{0.9}$Gd$_{0.1}$FeO$_3$ nanoprticles is 2.50 emu/g at 20 kOe for an average particle size of $11\pm 2$ nm and this is much higher compared to the values observed in Refs. \cite{ref16} and \cite{ref51}. The large value of magnetization may be associated with their reduced particle size as was observed in Ref. \cite{ref16} for single crystalline nultiferroic BiFeO$_3$ nanoparticles.  As was mentioned earlier, the magnetic ordering of BiFeO$_3$-based multiferroic ceramics is antiferromagnetic with a spiral modulated spin structure \cite{ref12, ref52, ref21} and an incommensurate long-wavelength period of 62 nm. The enhanced magnetic properties in these nanoparticles could be explained as : i) due to the modification of the long range spiral-modulated spin structure of BiFeO$_3$ the ferromagnetic properties were enhanced for the synthesized nanoparticles with sizes much smaller than 62 nm \cite{ref16}  ii) the ionic radius of Gd$^{3+}$ (0.938 {\AA}) ion is much smaller than that of Bi$^{3+}$ (1.17 {\AA}) ion which may also lead to large distortion in lattice structure thereby leading to reduction of spiral spin modulation in BiFeO$_3$ \cite{ref51} ; and  iii) Gd$^{3+}$ is magnetically active (effective magnetic moment is 8.0 Bohr Magnetron) and the ferromagnetic coupling between Gd$^{3+}$ and Fe$^{3+}$ ions may contribute significantly to enhance the magnetization \cite{ref51}.

In the case of bulk ceramics as well as nanoparticles, several authors reported that due to co-doping both in the Bi and Fe sites of BiFeO$_3$, multiferroic properties can be improved significantly \cite{ref12, ref54}. In a related system, Pb and Ti co-doped nanocrystalline Bi$_{0.9}$Pb$_{0.1}$Fe$_{0.9}$Ti$_{0.1}$O$_3$ multiferroics were prepared by solution combustion synthesis method from an $\alpha$ -alanine containing precursor \cite{ref54} for which the M$_s$ value was 0.60 emu/g at 40 kOe. In the investigation reported here, for 10 $\%$ Gd and Ti co-doped Bi$_{0.9}$Gd$_{0.1}$Fe$_{0.9}$Ti$_{0.1}$O$_3$ nanoparticles a well defined ferromagnetic hysteresis loop was observed in which the room temperature M$_s$ is 2.3 emu/g at 20 kOe. Both M$_s$ and M$_r$ of Bi$_{0.9}$Gd$_{0.1}$Fe$_{0.9}$Ti$_{0.1}$O$_3$ nanoparticles are much higher compared to that of sonochemically synthesized Bi$_{0.9}$Fe$_{0.95-x}$Sc$_{0.05}$Ti$_{x}$O$_3$ nanoparticles reported in a recent investigation \cite{ref92}.

In the case of bulk polycrystalline ceramics, due to the presence of magnetic secondary phases,  M$_s$, M$_r$ and H$_c$ were increased with the substitution of nonmagnetic Ti in place of Fe \cite{ref18}. On the contrary, in phase pure single crystalline nanoparticles, the substitution of nonmagnetic Ti reasonably decreased M$_s$, M$_r$ and H$_c$ although the net value of the saturation magnetization was always higher for each composition compared to that of bulk ceramics. The coercivity was found to be reduced for these nanoparticles compared to that of bulk ceramic samples as was also observed for the reduced size single crystalline BiFeO$_3$ nanoparticles \cite{ref16}. This indicates that the magnetic properties of these nanoparticles are related with their reduced size and crystallinity as was also determined in similar nanostructures \cite{ref16,ref51,ref54}. Compared to the bulk ceramics, the coercive fields of the synthesized nanoparticles are much smaller and almost negligible for Bi$_{0.9}$Gd$_{0.1}$Fe$_{1-x}$Ti$_x$O$_3$ (x= 0.20) nanoparticles (Table 1). This actually indicates their soft nature and better suitability for device applications \cite{ref71}.

\begin{table}[!h]
\caption{The table shows the comparison of the M$_{s}$, M$_{r}$ and  H$_{c}$ between Bi$_{0.9}$Gd$_{0.1}$Fe$_{1-x}$Ti$_x$O$_3$ nanoparticles and bulk ceramics \cite{ref18} observed at room temperature.}  
\begin{center}
\begin{tabular}{|l|l|l|l|l|l|l|}
 \hline
{Ti concentration} &\multicolumn{2}{c|}{M$_s$ (emu/g)}& \multicolumn{2}{c|}{M$_r$ (emu/g)} & \multicolumn{2}{c|}{H$_c$ (Oe)}\\
\cline{2-7}
 &Nano&Bulk&Nano&Bulk&Nano&Bulk \\
\hline
0.00&2.50&$-$&0.46&0.006&238&345\\
\hline
0.10&2.10&$-$&0.11&0.02&53&836\\
\hline
0.20&2.05&$-$&0.003&0.02&6.5&820\\
\hline
\hline
\end{tabular}
\end{center}
\end{table}

Notably, the perovskite manganites nanoparticles of various compositions were also synthesized directly from bulk powder using top-down approach based on ball milling \cite{ref98, ref56}. The longer milling time produced successfully nanoparticles from a few nanometers to hundred nanometers, however, the magnetic properties of the fine particles were found to degrade \cite{ref98}. This might be associated with the contamination and amorphorization of fine particles during the milling process \cite{ref57}. Although our material system is different, however, the findings of our investigation indicate the potentiality of the ultrasonication technique compared to mechanical milling for the preparation of phase pure nanoparticles which exhibit enhanced magnetic properties with size reduction.The product nanoparticles appear to have a good monodispersity with improved physical properties: the size, crystallinity and improved magnetic behavior of the synthesized nanoparticles at room temperature compared to the bulk starting material. 


\begin{figure}[hh]
\centering
\includegraphics[width=8.5cm]{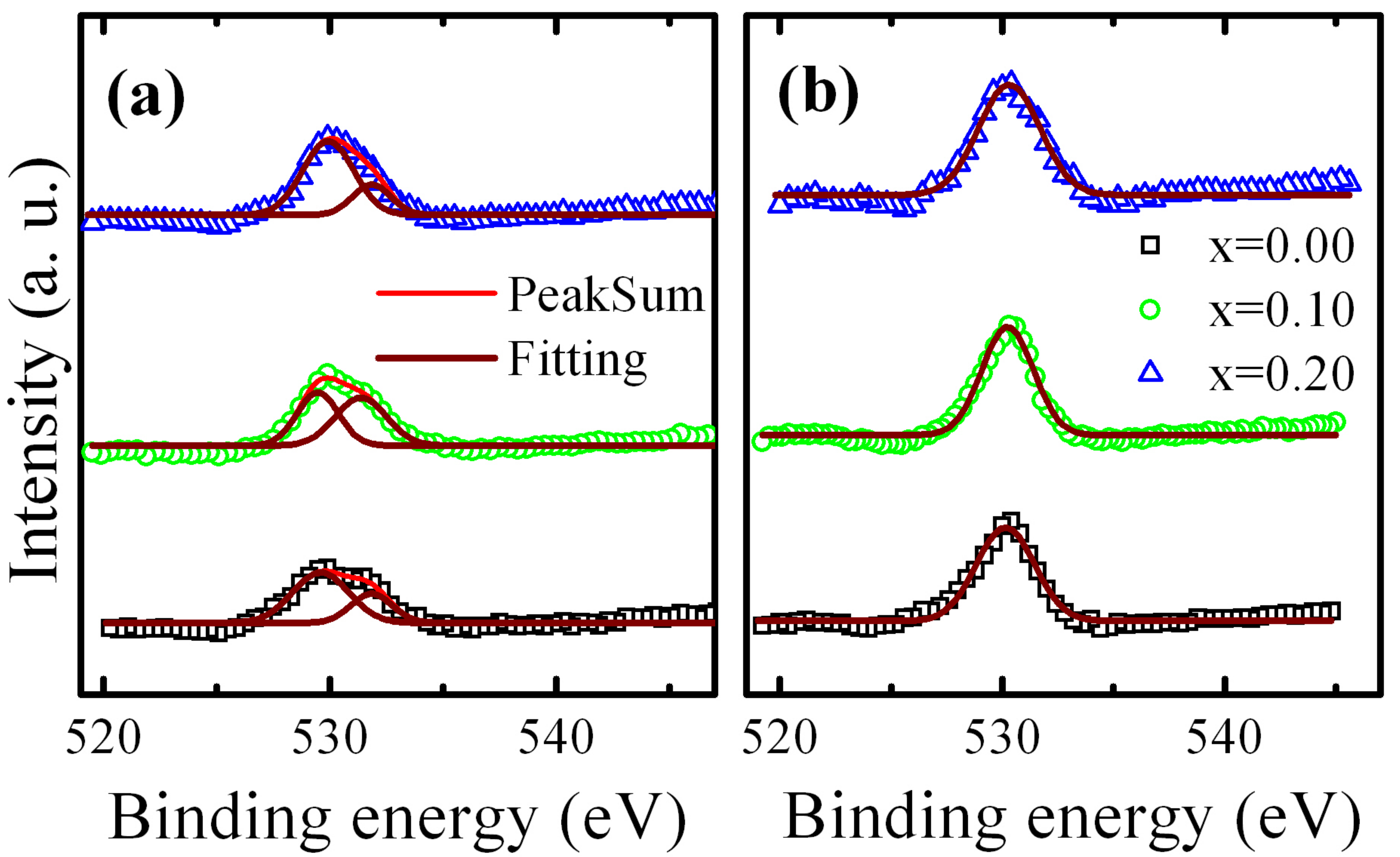}
\caption{XPS spectra of the O 1s of Bi$_{0.9}$Gd$_{0.1}$Fe$_{1-x}$Ti$_x$O$_3$ (x = 0-0.20) (a) bulk polycrystalline materials and (b) nanoparticles obtained after a sonication time of 60 minutes.} \label{fig6}
\end{figure} 

As was mentioned earlier, the bulk BiFeO$_3$ is characterized by serious current leakage problems due to the existence of a large number of charge centres caused by oxygen ion vacancies \cite{ref12, ref95} that degrades the ferroelectric properties. Therefore, to further confirm the concentration of oxygen vacancies in the bulk polycrystalline ceramics and nanoparticles of Bi$_{0.9}$Gd$_{0.1}$Fe$_{1-x}$Ti$_{x}$O$_3$ (x = 0.00, 0.10, 0.20)  XPS  measurements were also carried out.  As presented in Fig. \ref{fig6} (a), the O 1s XPS spectra of  bulk polycrystalline  Bi$_{0.9}$Gd$_{0.1}$Fe$_{1-x}$Ti$_{x}$O$_3$ (x = 0.00, 0.10, 0.20)  ceramics show a slightly asymmetric peak very close to 530 eV along with additional peak. The asymmetric curves of the bulk samples can be Gaussian fitted by two symmetrical peaks at 529.8 ev and  532 eV, respectively. The lower binding energy peak at 529.8 ev  is correspond the O 1s core spectrum, while higher binding energy peak is related to the oxygen vacancy \cite{ref63,ref93} in the sample. Interestingly, in the case of Bi$_{0.9}$Gd$_{0.1}$Fe$_{1-x}$Ti$_{x}$O$_3$ (x = 0.00, 0.10, 0.20)  nanoparticles we have observed a symmetrically single XPS peak (Fig. \ref{fig6} (b)) of O 1S at 530 eV \cite{ref92, ref94}. This indicates the absence of oxygen vacancy in Bi$_{0.9}$Gd$_{0.1}$Fe$_{1-x}$Ti$_{x}$O$_3$ (x = 0.00, 0.10, 0.20) nanoparticles. The absence of oxygen vacancies might reduce the leakage currents \cite{ref92} and therefore, we may expect improved ferroelectric properties of the synthesized nanoparticles in subsequent investigations.


\section{Conclusions} \label{II}
We have employed successfully a cost effective simple top-down approach to prepare ultrafine Bi$_{0.9}$Gd$_{0.1}$Fe$_{1-x}$Ti$_x$O$_3$ (x = 0.00-0.20) nanoparticles directly from bulk powder without using any chemical solution processes as well as toxic precursors. HRTEM imaging as well as electron diffraction techniques confirmed single crystal rhombohedral structure of these synthesized nanoparticles. Comparatively a long time of sonication of around 60 minutes allows fabricating ultrafine nanoparticles with narrow size distribution ranging from 11 nm to 13 nm. Magnetic measurements demonstrate the enhanced magnetic properties of the nanoparticles compared to that of bulk materials, which may be referred to as room temperature ferromagnetism caused by the size reduction. This method might be analogous to surfactant assisted ball milling creating magnetic nanoparticles \cite{ref4}, a technique which demonstrated a narrow size distribution of the synthesized particles with increasing milling time \cite{ref98, ref56}. The very narrow and small size distribution achieved in the present study are surprising when comparing to the broader and larger size distributions achieved in other sonofragmentations studies of other materials \cite{ref82,ref72,ref83,ref84}, but these results may first of all be material specific, and then the detection methods (SEM and various light scattering methods) employed in these previous studies are not always optimal to measure nanoparticles in the $\sim$ 10 nm size range found here. More detailed studies of the particle fragmentation would be relevant to include in future studies along with evaluation of the particles possible multiferroic properties by Polarization vs Electric field (P-E) hysteresis loop measurements \cite{ref51,ref79,ref94,ref20,ref92}. We conclude that this simple top-down preparation technique of ultrafine nanoparticles may be developed as a versatile technique for the preparation of other materials in general. 

\section{Acknowledgements}
University Grants Commission, Dhaka and the Bangladesh University of Engineering and Technology (BUET) are thanked for financial assistance and Department of Glass and Ceramic Engineering, BUET for some technical support.


\begin{thebibliography}{99}

\bibitem{ref1} W. Eerenstein, N. D. Mathur, J. F. Scott, Nature, 2006, \textbf{442}, 759.
\bibitem{ref2}	S. W. Cheong, M. Mostovoy, Nature Materials, 2007 \textbf{6}, 13-19.
\bibitem{ref3}	R. Ramesh, Nature, 461, \textbf{1218}, (2009).
\bibitem{ref53} K. F. Wang, J. M. Liu, Z. F. Ren,  Adv. Phys., 2009, \textbf{58}, 321-448.
\bibitem{ref77} J. T. Heron, D. G. Schlom, and R. Ramesh, App. Phys. Rev.  2014, \textbf{1}, 021303-18.
\bibitem{ref6}	V. V. Lazenka, G. Zhang, J. Vanacken, I. I. Makoed, A. F. Ravinski, V. V. Moshchalkov,  J. Phys. D: Appl. Phys., 2012, \textbf{45}, 125002-7.
\bibitem{ref7} M. S. Bernardo, T. Jardiel, M. F. Peiteado, J. Mompean, M. Garcia-Hernandez, M. A. Garcia, M.  Villegas, A. C. Caballero, Chem.  Mater., 2013, \textbf{25}, 1533-1541.
\bibitem{ref8} Q. H. Jiang, C. W.  Nan, J. Am. Ceram. Soc., 2006, \textbf{89}, 2123-2127.
\bibitem{ref9} S. Ghosh, A. S. Dasgupta, H. S. Maiti, J. Am. Ceram. Soc., 2005, \textbf{88}, 1349-1352.
\bibitem{ref10} T. Munoz, J. P. Rivera, A. Monnier, H. Schmid, Jpn. J. Appl. Phys. Part 1, 1985, \textbf{24}, 1051-1053.
\bibitem{ref11} M. Kumar, K. L. Yadav, J. Appl. Phys., 2006, \textbf{100}, 074111-4.
\bibitem{ref12}  R. A. Agarwal, S. S. Ashima, N. Ahlawat, J. Phys. D: Appl. Phys., 2012, \textbf{45}, 165001-9.
\bibitem{ref59} X. Qi, J. Dho, R. Tomov, M. G. Blamire, J. L. MacManus-Driscoll, App. Phys. Lett. 2005, \textbf{86}, 062903-3. 
\bibitem{ref52} P. Fischer, M. Polomska,  I. Sosnowska, M. Szymanski, J. Phys. C, 1980, \textbf{13}, 1931-1940.
\bibitem{ref13} G. Catalan, J. F.  Scott,  Adv. Mater., 2009, \textbf{21}, 2463-2485. 
\bibitem{ref14} I. Sosnowska, T. P. Neumaier, Steichele, J. Phys. C:
Solid State Phys. 1982, \textbf{15}, 4835-4846. 
\bibitem{ref15} C. Ederer, N. A. Spaldin, Phys. Rev. B, 2005, \textbf{71}, 060401(R)-4.
\bibitem{ref16} T. J. Park, Georgia C. A. Papaefthymiou, A. J. Viescas,
 A. R. Moodenbaugh, S. S. Wong, Nano Lett., 2007, \textbf{7} (3), 766-772.
\bibitem{ref17} A. Jaiswal, R. Das, K. Vivekanand, P. M. Abraham,  S. Adyanthaya, P. Poddar, J. Phys. Chem. C, 2010, \textbf{114}, 2108-2115.
\bibitem{ref51} W. Hu, Y. Chen, H. Yuan, G. Li, Y. Qiao, Y. Qin, S. Feng, J. Phys. Chem. C, 2011, \textbf{115}, 8869-8875.
\bibitem{ref75} F. Huang, Z. Wang, X. Lu, J. Zhang, K. Min, W. Lin, R. Ti, T. Xu,; J. He, C. Yue, J. Zhu, Scientific Reports, 2013, \textbf{3}, 2907. 
\bibitem{ref91} S. Gupta,  M. Tomar, V. Gupta, J. App. Phys., 2014, \textbf{115}, 014102-9. 
\bibitem{ref78} N. Nuraje, and K. Su, Nanoscale, 2013, \textbf{5}, 8752-8780.
\bibitem{ref79} M. M. Shirolkar, C. Hao,  X. Dong, T. Guo, L. Zhang, M. Li and H. Wang,  Nanoscale, 2014, \textbf{6}, 4735-4744. 
\bibitem{ref61} Y. Du, Z. X. Cheng,  X. L. Wang,  P. Liu,  S. X. Dou, J. App. Phys., 2011, \textbf{109}, 07B507-3.
\bibitem{ref81} J. L. O. Quinonez, D.  Diaz,  I. Z. Dube, H. A. Santamaria, I. Betancourt, P. S. Jacinto, N. N. Etzana, Inorganic Chemistry, 2013, \textbf{52} (18),10306-10316.
\bibitem{ref63} L. A. Fang, J. A. Liu, S. Ju,  F. G. Zheng, W.  Dong, M. R. Shen, Appl. Phys. Lett., 2010, \textbf{97}, 242501-3.
\bibitem{ref94} D. P. Dutta, B. P. Mandal, R. Naik, G.  Lawes, A. K. Tyagi, J. Phys. Chem. C, 2013, \textbf{112}, 2382-2389.
\bibitem{ref96} I. Hernández-Perez, A. M. Maubert, Luis Rendón, Patricia Santiago, H. Herrera-Hernández,L. Díaz-Barriga Arceo, V. Garibay Febles, Eduardo Palacios González, L. González-ReyesInt. J. Electrochem. Sci.,2012, \textbf{112}, 8832-8847.
\bibitem{ref82} K. R. Gopi and R. Nagarajan, IEEE Transactions on Nanotechnology, 2008, \textbf{7(5)} 532-7.
\bibitem{ref83} K. A. Kusters, S. E. Pratsinisap, S. G. Thorna, D. M. Smith, Powder Technology. 1994, \textbf{80(3)}, 253-63. 
\bibitem{ref84} M. D. Kass, Materials Letters, 2000, \textbf{42(4)} 246-50. 
\bibitem{ref85} Y. Y. Huang, T. P. J. Knowles, and E. M. Terentjev, Advanced Materials, 2009 \textbf{21(38-39)}, 3945-8. 
\bibitem{ref18} M. A. Basith,  O. Kurni,  M. S. Alam,  B. L. Sinha, B. Ahmmad, J. Appl. Phys., 2014, \textbf{115}, 024102-7.
\bibitem{ref19} I. O. Troyanchuk, A. N. Chobot, O. S. Mantytskaya, N. V. Tereshko, Inorg. Mater., 2010, \textbf{46}, 424-428.
\bibitem{ref20} G. S. Lotey, N. K. Verma, J. Nanopart. Res. 2012, \textbf{14}, 742-748.
\bibitem{ref86} Z. R. Dai, S. Sun, Z. L. Wang, Surf. Sci., 2002, \textbf{505}, 325.
\bibitem{ref87} J. Sort, S. Surinach, S. Munoz, M. D. Baro, M. Wojcik, E. Jedryka, S. Nadolski, N. Sheludko, and J. Nogues, Phys. Rev. B, 2003, \textbf{68}, 014421.
\bibitem{ref88} A. Recnik, I. Nyiõ-Kósa, I. Dódony, M. Pósfai, CrystEngComm, 2013, \textbf{15}, 7539.
\bibitem{ref97} J. S. Taurozzi, V.A. Hackley, M. R. Wiesner, Nanotoxicology, 2011 \textbf{5(4)}, 711–29.
\bibitem{ref55} J. J. Neumeier, H. Terashita, Phys. Rev. B., 2004, \textbf{70}, 214435-7.
\bibitem{ref21} D. Lebeugle, D. Colson, A. Forget, M. Viret, P. Bonville, J. F. Marucco, S. Fusil, Phys. Rev. B., 2007, \textbf{76}, 024116-8.
\bibitem{ref58} R. Guo, L. Fang, W. Dong, F. Zheng, M. Shen, J. Phys. Chem. C, 2010, \textbf{114}, 21390-21396.
\bibitem{ref54} K. Singh, R. K. Kotnala, M. Singh, App. Phys. Lett.  2008, \textbf{93}, 212902-3.
\bibitem{ref92} D. P. Dutta, B. P. Mandal, M. D. Mukadam, S. M. Yusuf, A. K. Tyagi, Dalton Trans., 2014,  \textbf{43}, 7838-7846.
\bibitem{ref71} S. R. Shannigrahi, A. Huang, N. Chandrasekhar, D. Tripathy,
A. O. Adeyeye,  Appl. Phys. Lett. 2007, 90, 022901.
\bibitem{ref98} The-Long Phan, New Physics: Sae Mulli, 2013, \textbf{63}, 557-561.
\bibitem{ref56} S. Roy, I. Dubenko, D. D. Edorh, and N. Ali, J. Appl. Phys. 2004 \textbf{96} 1202-1208. 
\bibitem{ref57} V. M. Chakka, B. Altuncevahir, Z. Q. Jin, Y. Li and J. P. Liu, J. Appl. Phys. 2006 \textbf{99} 08E912. 
\bibitem{ref95} A.R. Makhdoom, M. J. Akhtar, M. A. Rafiq, M. M. Hassan, Ceramics International, 2012, \textbf{38}, 3829-3834. 
\bibitem{ref93} R. Das, T. Sarkar, K. Mandal, J. Phys. D: Appl. Phys. 2012, \textbf{45}, 455002-12.
\bibitem{ref4}	 Y. Wang, Y. Li, C. Rong and J. P. Liu, Nanotechnology, 2007, \textbf{18}, 465701.
\bibitem{ref72} U. Teipel, K. Leisinger, I. Mikonsaari, International Journal of Mineral Processing. 2004 \textbf{74} S183-S190. 






\end{thebibliography}
\end{document}